# Superoscillation focusing of high-order cylindrical-vector beams

Zhongwei Jin,[1,2,4] Yijie Jin,[1,3] Fangzhou Shu,[2,5] Bin Fang,[1] Zhi Hong,[2] Jianjun Liu,[2] Yuhang Yao,[1] Keyi Chen,[1] Shengtao Mei,[1,6]

[1] College of Optical and Electronic Technology, Centre for THz Research, China Jiliang University, Hangzhou 310018, China
[2] Centre for THz Research, China Jiliang University, Hangzhou 310018, China
[3] Artificial Intelligence Innovation Center, Yangtze Delta Region Institute of Tsinghua University, Zhejiang 314006, China
[4] jinzhongwei@cjlu.edu.cn

[5] fzshu@cjlu.edu.cn

[6] meishengtao@gmail.com

**Traditional superoscillation focusing typically requires complex optimization of the incident light field. These complexities may limit the practical application of superoscillation. High-order radially polarized Laguerre-Gaussian beams inherently support superoscillation focusing due to their multi-ring amplitude distribution and 0-$\pi$ phase alternation, which align with the necessary destructive interference mechanisms. In this study, we demonstrate that by adjusting the beam's mode order together with the incident beam size, we can easily control the full width at half maximum, field of view, and energy distribution of superoscillation focusing. Moreover, high-order azimuthally polarized vortex-phase Laguerre-Gaussian beams can also achieve superoscillation focusing, offering even better super-resolution effects. The distinct focusing behaviors of their circular components present unique opportunities for applications involving circular dichroism materials.**

Super-resolution imaging techniques have always been a powerful means for people to continuously explore the microscopic world, and reducing the full width at half maximum (FWHM) of the point spread function (PSF) of the imaging system is the most direct way to achieve super-resolution imaging[1, 2]. Stimulated Emission Depletion (STED) microscopy is one of the most successfully commercialized super-resolution techniques to date[3]. Additionally, subtraction imaging technology achieves an equivalent effect through two imaging processes[4]. Whether introducing additional materials or performing multiple imaging processes, both will complicate the imaging process to some extent and may even cause unavoidable damage to the imaged object. Therefore, directly constructing a PSF that breaks through the traditional diffraction limit may be the most advantageous solution.

Superoscillation focusing achieves subwavelength focusing by utilizing the property of superoscillatory functions, which can oscillate faster in a local region than the highest frequency component of their Fourier spectrum [5-7]. However, previous studies have shown that while superoscillation focusing can surpass the diffraction limit, it comes at the cost of low energy efficiency and unavoidable high-energy side lobes[8, 9]. Furthermore, whether you are using a metasurface with a complex amplitude distribution [10], constructing an aperture filter with a specific light field distribution in front of the objective lens[7], or employing a superoscillatory lens that combines light field modulation and focusing functions[7], it is essential to carefully design the complex field distribution at the light field's input. In recent years, researchers have discovered that higher-order radially polarized Laguerre-Gaussian (RP-LG$_{p,1}$) beams have inherent advantages in achieving superoscillation focusing[11].

In this work, we demonstrate the effects of the mode order ($p$) and polarization state of cylindrical-vector (CV) LG beams on their superoscillation focusing. We find that higher-order LG beams can push the side lobes around the superoscillation focal point towards the periphery, thereby achieving a larger field of view (FoV). In addition to controlling the size of the incident beam, we can also achieve fine-tuning of superoscillation focusing by controlling the beam's mode order ($p$). Furthermore, we discover that higher-order azimuthally polarized vortex-phase Laguerre-Gaussian (APV-LG) beams can also achieve superoscillation focusing, they exhibit even better super-resolution performance under the same conditions. The focusing differences between their left- and right-handed components hold promise for unique applications in scenarios involving chiral materials.

In the scalar paraxial case, when focusing a circular aperture by a lens, produces what we commonly refer to as an Airy disk, which has the form $|J_1(v)/v|^2$ ($v = kr$NA, $k$: wavenumber in vacuum, $r$: radial position, NA: numerical aperture, $J_1$ is the first-order Bessel function of the first kind) and its size is regarded as the Rayleigh criterion (RC)[2, 12]. Equivalently, it can be expressed as $|J_0(v) + J_2(v)|^2$. The latter expression might be more referential because, under the same conditions, an infinitely narrow annular aperture masked beam focused by a lens yields a standard Bessel beam $|J_0(v)|^2$.

Clearly, in the scalar paraxial case, light from a narrow annular aperture can be focused more tightly than that from a circular aperture. Considering only the light's highest angular frequency transmission through the lens, we can obtain the minimum focal spot size in the scalar paraxial focusing case as $0.359\lambda_0/\text{NA}$ ($\lambda_0$ is the wavelength in vacuum)[13]. Further considering the focusing of a narrow annular aperture by a high-NA objective lens, only linear polarization, circular polarization, and radial polarization can focus on the propagation axis. All of these polarizations introduce higher-order Bessel functions ($J_1$ & $J_2$) at the focal point[12]. Specifically, for linear and circular polarizations, the energy proportion of higher-order Bessel field increases with increasing NA, thereby broadening the focused spot. The focusing distribution of RP light is $\text{NA}^2/n^2|J_0(v)|^2+(1-\text{NA}^2/n^2)|J_1(v)|^2$ (n is the refractive index of the medium), and it is evident that as NA increases, the $J_0$ component becomes increasingly dominant, and the energy distribution gradually approaches the focusing field distribution of an infinitely narrow annular aperture in the scalar paraxial case. In this scenario, we can obtain the minimum focal spot size of RP light focused by an objective lens after passing through an infinitely narrow annular aperture as $0.379\lambda_0/\text{NA}$ [7, 11].

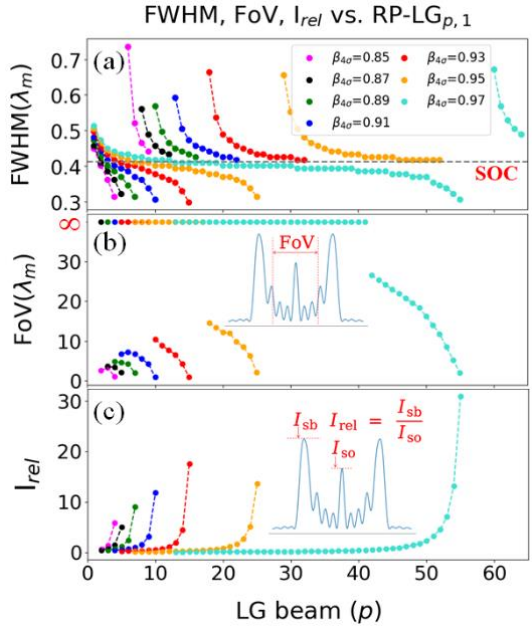

**Fig.1.** Superoscillation focusing properties regarding LG beam's mode order. (a) Focal spot's FWHM as a function of mode order ($p$) with selected beam size parameters ($\beta_{4\sigma}$ = 0.85, 0.87, 0.89, 0.91, 0.93, 0.95, 0.97), (b) Focal plane's FoV when superoscillation focusing happening. The maximum value on the $y$-axis is infinity, corresponding to cases where the center focal spot's intensity is sufficiently large. (c) Relative intensity $I_{rel}$ of the maximum side lobe on the superoscillation focal plane when increasing mode order.

By dividing a higher-order RP-LG beam into inner rings and an outermost ring, and then adjusting the overall beam size by applying a cutoff aperture to the energy of the outermost ring, the original constructive interference condition between the inner and outer ring fields is inevitably disrupted. Continuously adjusting the size of the cutoff aperture can ultimately achieve destructive interference, leading to superoscillation focusing, which is not surprising. However, the reported lower-order RP-LG beams in superoscillation focusing scenarios exhibit sensitivity of side lobe energy to beam size, especially when the superoscillation phenomenon is pronounced (e.g., when FWHM < $0.4\lambda_m$, $\lambda_m = \lambda_0/n$)[11].

Here we first consider the mode order ($p$) influence on RP-LG$_{p,1}$ beams in superoscillation focusing, which is not systematically studied previously. In our calculations, NA=1.4, n=1.52 and $\lambda_0$=532nm. The gray dashed line in Figure 1(a) represents the FWHM of the focal point corresponding to the highest angular frequency of the RP beam $0.412\lambda_m$), which is also the FWHM threshold for its superoscillation focusing. When the incident beam size is fixed, continuously adjusting the mode order of the beam ($p$=1~64) allows the size of the focused spot to exhibit a quasi-hyperbolic trend between diffraction-limited and superoscillation focusing. When superoscillation focusing occurs (i.e., the corresponding points fall below the dashed line), we plot the FoV of the superoscillation focusing field and the relative intensity ratio $I_{rel}$ in Figures 1(b)(c)[8]. When the beam size parameter $\beta_{4\sigma}$ is 0.87 or above, the superoscillation focal intensity of lower-order RP-LG beams can be several times that of the strongest side lobe. At this point, FoV is infinitely large, and $I_{rel}$ is very small. As the $\beta_{4\sigma}$ increases to 0.95 and above, we can achieve both superoscillation focusing and strong focal intensity over a wide range of mode orders, which is quite rare for superoscillation focusing. Of course, this property is present only when the superoscillation focal spot size is slightly smaller than that of the highest angular frequency (corresponding to cases that FoV reaching infinity as shown in Figure 1(b)). As the focal spot size further decreases, the FoV gradually shrinks, and the $I_{rel}$ of the side lobes further increases. When the focal spot size approaches $0.3\lambda_m$, the FoV shrinks to 1~2$\lambda_m$, and $I_{rel}$ can be 20 or even 30. Even the intensity of the nearest side lobes can exceed that of the superoscillation focal point. In this scenario, further reducing the focal spot size becomes practically meaningless for applications, as even a 0.5 Airy unit (AU) pinhole in confocal laser scanning microscope (CLSM) would struggle to isolate the central focal point for imaging.

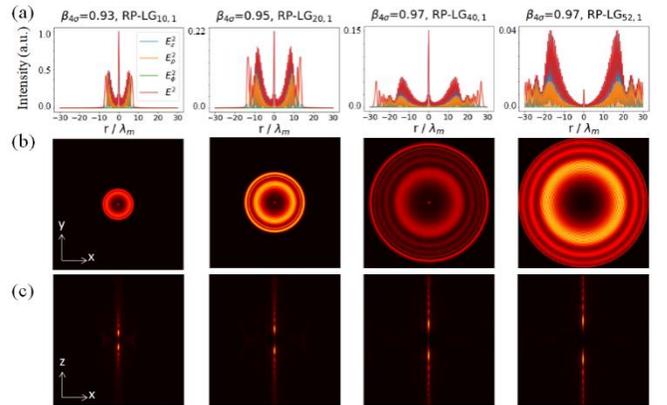

**Fig.2.** Superoscillation focusing with selected mode order $p$=10, 20, 40, 52 and appropriate beam parameters. (a) Radial intensity profiles of the focus with four different ($\beta_{4\sigma}$, $p$) settings. (b) The corresponding $xy$ cross-

section intensity distributions on the focal plane. (c) The corresponding xz cross-section intensity distributions. The focal planes lie in the center of the on-axis intensity discontinuities.

In Figure 2, we show the superoscillation focusing phenomena of several higher-order RP-LG$_{p,1}$ beams with parameters (($\beta_{4\sigma}$, $p$) = (0.93, 10), (0.95, 20), (0.97, 40), (0.97, 52)). As $p$ increases from 10 to 40, their FWHM remains similar (~0.38$\lambda_m$), with focal intensities all higher than the strongest side lobe (Figure 2(a)). The energy distribution in the $xy$ cross-section of the focal point shows that more energy is pushed further away from the center (Figure 2(b)). In the $xz$ plane, there are increasingly longer intervals of discontinuity in the intensity along the central axis before and after the focal plane. This is because, with $p$ increasing, more energy is pushed further away from the central axis (Figure 2(c)). In the fourth column of Figure 2, we select the combination (0.97, 52), where the corresponding focal spot FWHM is approximately 0.346$\lambda_m$, with its FoV exceeding 10$\lambda_m$. Although most of the energy is concentrated in distant side lobes, the focal intensity is significantly higher than that of the nearby side lobes, making it easily filterable through a pinhole in CLSM. Combining Figures 1&2, it can be observed that when RP-LG beams achieve superoscillation focal spots of similar size at different $p$, higher-order beams exhibit a better signal-to-noise ratio for the focal spot intensity relative to nearby side lobe intensity. This suggests potentially improved performance in subsequent imaging or other applications.

AP light cannot converge into a single focal point when focused [14]. However, researchers discovered that by applying an additional vortex phase to AP light, it can not only achieve focusing but potentially outperform the RP light[15]. Previous numerical simulations and experiments have demonstrated the excellent performance of APV beam in achieving lateral super-resolution and ultra-long working distances axially[16]. These results suggest that APV beam may also have potential in superoscillation focusing. We begin by considering the focusing of APV beam and examine the scenario where an infinitely narrow annular aperture is placed in front of the lens. In this case, the electric field near the focal point can be expressed as

$$E(\rho, \varphi, z) = \frac{1}{2} kfA \sqrt{\cos\theta_0} \sin\theta_0 \exp(ikz\cos\theta_0),$$

where

$$A = \begin{bmatrix} e^{2i\varphi} J_2(k\rho\sin\theta_0) + J_0(k\rho\sin\theta_0) \\ -i\left(e^{2i\varphi} J_2(k\rho\sin\theta_0)\right) - J_0(k\rho\sin\theta_0) \\ 0 \end{bmatrix} \quad (1)$$

Equation (1) presents the expressions for the three electric field components ($E_X$, $E_Y$, $E_Z$) near the focal point[17]. Clearly, after focusing, the AP light with a vortex phase still only has $x$ and $y$ components. Through a simple transformation, we can convert the above electric field into a form based on circular polarization ($E_L$, $E_R$, $E_Z$):

$$A = \frac{\sqrt{2}}{2} \begin{bmatrix} J_0(k\rho\sin\theta_0) \\ e^{2i\varphi} J_2(k\rho\sin\theta_0) \\ 0 \end{bmatrix} \quad (2)$$

Evidently, Equation (2) has a more concise form and its physical meaning is clearer: the left-handed circularly polarized light $E_L$ at the focal point has the form $J_0$, while the right-handed circularly polarized light $E_R$ around the focal point has the form $J_2$, and carries a vortex phase of order 2. It is worth emphasizing here that the left-handed polarized light focused on the axis is a result of spin-orbit interaction, while the off-axis right-handed polarized light similarly adheres to the principle of angular momentum conservation. Consequently, the energy distribution on the focal plane can be written as:

$$\frac{1}{2}\left(\left|J_0(v)\right|^2 + \left|J_2(v)\right|^2\right) \quad (3)$$

This form is similar to the energy distribution at the focal point of RP light. The energy distribution of left-handed circularly polarized light ($E_L$) is consistent with that of RP light's longitudinally polarized part ($E_Z$) at focus, and the energy distribution of right-handed circularly polarized light ($E_R$) matches the transverse energy distribution of RP light ($E_\rho$). The main difference is that in this formula, the ratio of $J_0$ to $J_2$ is independent of lens parameters, with the energy ratio of the two orthogonal components fixed at 1:1. This may not be advantageous for focusing, as we would likely prefer $J_0$ to dominate, but there might be other applications requiring such a property, which is beyond the scope of discussion here. The superoscillation focusing criterion of APV beam from (3) is 0.400$\lambda_m$, which is smaller than that of RP beam (0.412$\lambda_m$).

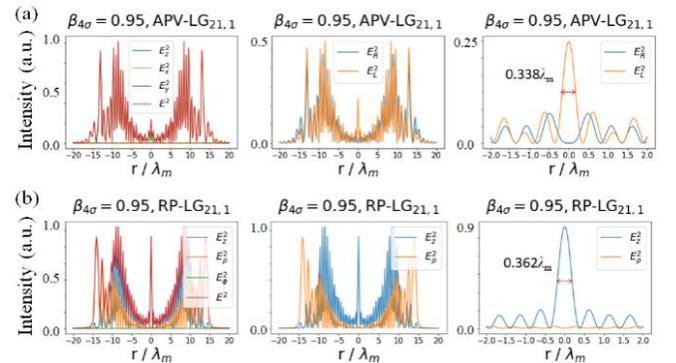

**Fig.3.** Comparison of APV-LG and RP-LG superoscillation focusing. (a) APV-LG$_{21,1}$ radial intensity profile at the focus for different components. Superoscillation focusing of APV-LG mainly happens on its $E_L$ part, while the $E_R$ part forms a hollow shape sharing roughly the same energy. (b) RP-LG$_{21,1}$ intensity profile at the focus. The main contribution of focusing is from $E_Z$ and lateral field is pushed away to form side lobes at the periphery.

We found similar conclusions can be drawn for APV-LG$_{p,1}$ under the same focusing conditions with RP-LG$_{p,1}$ as shown in Figure 1. Therefore, in Figure 3, we present the focusing scenarios of RP-LG and APV-LG beams at (0.95, 21). In terms of overall superoscillation focusing effect, APV-LG beams outperform RP-LG beams with smaller FWHM (0.354$\lambda_m$ < 0.37$\lambda_m$). Regarding the total energy distribution, focusing of RP-LG achieves higher focal spot intensity and significantly larger FoV, while APV-LG beams have relatively stronger-intensity side lobes near the main focal spot. We can observe that the main components of on-axis focusing for both CV beams ($E_L$ & $E_Z$) have very similar form of $J_0$ (FWHM: $0.338\lambda_m$ ($E_L$) < $0.362\lambda_m$ ($E_Z$)), while there is a larger energy gap in the other two components ($E_R$ and $E_\rho$), where RP-LG shows its advantage. This is the fundamental reason why APV beam's focal point is always accompanied by a stronger side lobe.

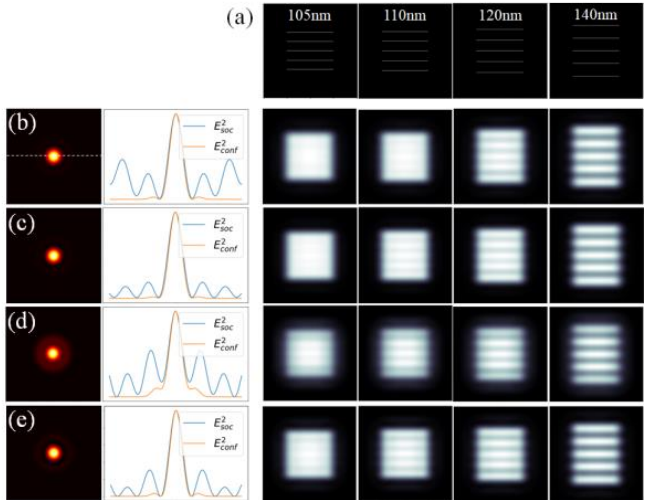

**Fig.4.** Numerical simulations of LG beam assisted CLSM imaging. (a) Simulated line objects with spacing of 105nm, 110nm, 120nm and 140nm. (b) Left: RP-LG$_{3,1}$ confocal PSF $xy$ distribution and radial intensity profiles along the white dash line ($E_{soc}$ is the excitation source intensity profile and $E_{conf}$ is the confocal system intensity profile), Right: simulated images of different spacing images. (c, d) Results of RP-LG$_{21,1}$, APV-LG$_{21,1}$. (e) Results of when only considering the $E_L$ part of APV-LG$_{21,1}$ as the excitation signal.

To further compare the superoscillation focusing effects of higher-order CV LG beams, we simulated the phenomenon of using these beams for CLSM imaging in Figure 4. We use a 532 nm laser as the excitation source, with the excited fluorescence wavelength set to 1.05 times that of the excitation source. We use a 100x magnification objective lens system with NA=1.4 (n=1.52) and a pinhole of 0.5 AU. We assume line-shaped objects for imaging tests (as shown in Figure 4(a)). First, we compare the imaging effects for RP-LG cases (0.85, 3) and (0.95, 21). The superoscillation focal FWHM of the excitation source is $0.37\lambda_m$ for both settings. The difference is that at $p=21$, the focusing has a larger FoV and smaller adjacent side lobe intensity. Although an AU=0.5 pinhole can effectively filter out the side lobes, it can still be observed that the confocal system's PSF at $p=3$ has slightly stronger side lobe intensity (see Figures 4(b) and (c)). Consequently, when the line object spacing (110 nm) is less than the system's confocal PSF FWHM, the imaging resolution at $p=21$ is slightly better than at $p=3$. APV-LG at (0.95, 21) provides the system with higher resolution (confocal PSF FWHM of 114 nm), and it still has some resolution capability even for objects with a 105 nm spacing. Comparing Figures 4(c) and 4(d), the APV-LG beam in superoscillation focusing imaging outperforms the RP-LG under the same conditions. Moreover, since the former has a transverse field at the focal plane, while the latter is dominated by a longitudinal field, when the light field around the focal point passes through an interface with a refractive index change (e.g., from air to glass), APV-LG beam can maintain its focal spot size, while RP-LG beam's focus will significantly broaden. We would like to further emphasize here that, by using some chiral materials, such as chiral plasmonic materials or chiral molecular materials, we can directly filter out the left-handed light energy undergoing superoscillation focusing on the focal plane, achieving a narrower excitation light FWHM. Alternatively, a simpler approach in a transmission-type CLSM system is to use a wave plate to observe only the left-handed light signal, thereby achieving better resolution with virtually no additional cost.

In conclusion, we have analyzed the patterns and differences in superoscillation focusing of higher-order CV LG beams. Building upon previous studies that focused on adjusting the incident beam size of RP-LG beams, we have further demonstrated the impact of mode order on superoscillation focusing. More importantly, we discovered that APV-LG beams can also achieve superoscillation focusing. Under the same incident beam size and mode order, APV-LG beams can achieve a smaller superoscillation focal point and exhibit a unique left- and right-handed polarization separation phenomenon near the focal point. Based on these phenomena and the simulation results, the superoscillation focusing of higher-order CV LG beams can achieve higher resolution in CLSM imaging and may be beneficial for applications in other related areas.

**Funding.** National Natural Science Foundation of China (12304434, 12004362, 12204446).

**Disclosures.** The authors declare no conflicts of interest.

**Data availability.** The data that support the findings of this study are available within the article and from the corresponding author upon reasonable request.